\documentstyle[amsfonts,prl,aps,epsfig,twocolumn]{revtex}

\begin{document}
\draft

\twocolumn[\hsize\textwidth\columnwidth\hsize\csname@twocolumnfalse\endcsname
\title{Quantum state transfer via the ferromagnetic chain in a spatially modulated field }
\author{T. Shi $^{1}$,Ying Li $^{1}$, Z. Song$^{1,a}$, and C. P. Sun $^{1,2,b,c}$}
\address{ $^{1}$Department of Physics, Nankai University, Tianjin
300071, China\\
$^{2}$ Institute of Theoretical Physics, The Chinese Academy of
Science, Beijing, 100080, China\\
}
 \maketitle

\begin{abstract}

We show that a perfect quantum state transmission can be realized
through  a   spin chain possessing   a commensurate structure of
energy spectrum, which is matched with the corresponding parity.
As an exposition of the mirror inversion symmetry discovered by
Albanese et. al (quant-ph/0405029), the parity matched the
commensurability of energy spectra help us to present the novel
pre-engineered spin systems for quantum information transmission.
Based on the these theoretical analysis, we propose a protocol of
near-perfect quantum state transfer by using a ferromagnetic
Heisenberg chain with uniform coupling constant, but an external
parabolic magnetic field. The numerical results shows that the
initial Gaussian wave packet in this system with optimal field
distribution can be reshaped near-perfectly over a longer
distance.

\end{abstract}

\pacs{PACS number: 03.65.-w, 03.67.Lx, 42.50.Gy, 31.15.Lc} ]

\bigskip

Recently the quantum information processing (QIP)\ protocols have
been considered with the quantum spin \cite{Div,cirac2,cirac1} (or
quasi-spin \cite{s-prl}) systems. The simple spin chains have been
explored as a coherent data bus \cite{Bose,Subra,Matt}. It
provides us with a quantum channel for perfect transmission of
quantum state when the spin chain is pre-engineered \cite{Perfc},
an isotropic antiferromagnetic spin ladder system was proposed as
a novel robust kind of quantum data bus \cite{songz}. Due to a
large spin gap existing in such a perfect medium, the effective
Hamiltonian of the two connected spins can be archived as that of
Heisenberg type by adiabatic elimination, which possesses an
effective coupling strength inversely proportional to the distance
of the two spins and thus the quantum information can be
transferred between the two spins separated by a longer distance,
i.e. the characteristic time of quantum state transferring
linearly depends on the distance. Such a gapped spin system can be
used as a perfect quantum channel for perfect quantum state
transmission if local measurements on the individual spins can be
implemented. In fact, it has been proved by F. Verstraete et al
\cite{cirac2,cirac1} that the ground states of a spin system with
energy gap possesses an infinite entanglement length opposed to
their finite correlation length.

The physical process of quantum state transmission through a
quantum spin system can be understood as an dynamical permutation
(or a translation) preserving the initial shape of a quantum state
of the involved two qubits, which can be realized as an specific
evolution of the total quantum spin system from an initial wave
function localized a single site of the lattice to another in long
distance. Most recently it was discovered that, if there exists a
mirror inversion symmetry (MIS) with respect to its center in the
spin chain, such quantum evolution can occur dynamically at
certain instant \cite{MIS}. Such scheme for quantum state
transmission is much appreciated since no dynamical control is
required over individual qubits. In this article we will revisit
this elegant conception by explicitly considering the spectrum
structure and the corresponding parities of such MIS system. We
discover that the MIS can be implemented in a universal quantum
spin system with a commensurate spectral structure matching with
the corresponding parities. With the help of this discovery, we,
in principle, can propose various novel scenarios for perfect and
near-perfect quantum information transmission through the
pre-engineered quantum spin chains. We give an example seems to be
more complicated than those by others.

Furthermore, a scheme, based on our theoretical analysis to
realize near-perfect quantum state transfer, is proposed with the
quantum channel by a ferromagnetic Heisenberg chain with uniform
coupling constant, but an external parabolic magnetic field .
Numerical results show that, for the optimal field distribution,
this system can perform a near-perfect transfer for a Gaussian
wave packet over a longer distance.

To sketch our central idea, let us first consider a single particle system
with the usual spatial refection symmetry (SRS) in the Hamiltonian $H$. Let $%
P$ be the spatial refection operator. The SRS is implied by
$[H,P]=0$.
Now we prove that, after time $\pi /E_{0}$ any state $\psi (\overrightarrow{r%
})$ can evolve into the reflected state $\pm \psi
(-\overrightarrow{r})$ if the eigenvalues $\varepsilon _{n}$ match
the parities $p_{n}$ in the following way
\begin{equation}
\varepsilon _{n}=N_{n}E_{0}, p_{n}=\pm (-1)^{N_{n}}
\end{equation}%
for the arbitrary positive integer $N_{n}$ and

\begin{equation}
H\phi _{n}(\overrightarrow{r})=\varepsilon _{n}\phi _{n}(\overrightarrow{r}%
), P\phi _{n}(\overrightarrow{r})=p_{n}\phi
_{n}(\overrightarrow{r}).
\end{equation}%
Here, $\phi _{n}(\overrightarrow{r})$ is the common eigen wave function of $%
H $ and $P$, $\overrightarrow{r}$ the position of the particle. We
call Eq (1) the spectrum-parity matching condition (SPMC).

The proof of the above rigorous conclusion is a simple, but heuristic
exercise in basic quantum mechanics. In fact, for the spatial refection
operator, $P\psi (\overrightarrow{r})=\pm \psi (-\overrightarrow{r})$. For
an arbitrarily given state at $t=0,$ $\psi (\overrightarrow{r},t)\left\vert
_{t=0}\right. =\psi (\overrightarrow{r})$, this state evolves to
\begin{equation}
\psi (\overrightarrow{r},t)=e^{-iHt}\psi (\overrightarrow{r}%
)=\sum_{n}C_{n}e^{-iN_{n}E_{0}t}\phi _{n}(\overrightarrow{r})
\end{equation}%
at time $t$, where $C_{n}=\left\langle \phi _{n}\right\vert \psi \rangle$.
Then, at time $t=\pi /E_{0}$, we have

\begin{eqnarray}
\psi (\overrightarrow{r},\pi /E_{0})
&=&\sum_{n}C_{n}(-1)^{N_{n}}\phi _{n}(%
\overrightarrow{r})   \\
&=&\pm \sum_{n}C_{n}p_{n}\phi _{n}(\overrightarrow{r})\nonumber \\
&=&\pm P\psi (%
\overrightarrow{r})=\pm \psi (-\overrightarrow{r}).\nonumber
\end{eqnarray}%
This is just the central result \cite{MIS} discovered for quantum spin
system that the evolution operator can become a parity operators $\pm P$ at
some instant $t=(2n+1)\pi /E_{0}$, that is

\begin{equation}
\exp [-iH\pi /E_{0}]=\pm P.
\end{equation}

From the above arguments we have a consequence that if the eigenvalues $%
\varepsilon _{n}=N_{n}E_{0}$ of a 1-D Hamiltonian $H$ with spatial refection
symmetry are odd-number spaced, i.e. $N_{n}-N_{n-1}$ are always odd, any
initial state $\psi (x)$ can evolve into $\pm \psi (-x)$ at time $t=\pi
/E_{0}$. In fact, for such 1-D system, the discrete states alternate between
even and odd parity. Consider the eigenvalues $\varepsilon _{n}=N_{n}E_{0}$
are odd-number spaced, the next nearest level must be even-number spaced,
then the SPMC are satisfied. Obviously, the 1-D SPMC is more realizable for
the construction of the model Hamiltonian to perform perfect state transfer.

Now, we can directly generalize the above analysis to many particle systems.
\ For the quantum spin chain, one can identify the above SRS as the MIS with
respect to the center of the quantum spin chain. As the discussion in ref.%
\cite{MIS}, we write MIS operation
\begin{equation}
P\Psi (s_{1,}s_{2,}...,s_{N-1,}s_{N})=\Psi (s_{N},s_{N-1,}...,s_{2},s_{1,})
\end{equation}%
for the wave function $\Psi (s_{1,}s_{2,}...,s_{N-1,}s_{N})$ of
spin chain. Here, $s_{n}=0,1$ denotes the spin values of the n-th
qubit. According to this representation of SRS and our discovered
SPMC, many spin systems can be pre-engineered for perfect quantum
states transfer. For instance, two-site spin-$\frac{1}{2}$
Heisenberg system the simplest example which meets the SPMC.
Recently, M. Christandl et al \cite{Matt} proposed a $%
N$-site $XY$ chain with an elaborately designed modulated coupling
constants between two nearest neighbor sites, which ensures a
perfect state transfer. It is easy to find that this model
corresponds the SPMC for the simplest case $N_{n}=n$.

In the following, we propose a class of different models for
perfect state transfer, whose spectrum structures obey our SPMC
exactly. Consider a $N$-site spin$-\frac{1}{2}$ $XY$ chain with
the Hamiltonian

\begin{equation}
H=2%
\mathop{\textstyle\sum}%
_{i=1}^{N-1}J_{i}[S_{i}^{x}S_{i+1}^{x}+S_{i}^{y}S_{i+1}^{y}]
\end{equation}%
where $S_{i}^{x},S_{i}^{y}$ and $S_{i}^{z}$ are Pauli matrices for the $i-$%
th site, $J_{i}$ \ the coupling strength for nearneibour interaction. For
the open boundary condition, this model is equivalent to the spin-less
fermion model. The equivalent Hamiltonian can be written as

\begin{equation}
H=%
\mathop{\displaystyle\sum}%
\limits_{i=1}^{N-1}J_{i}^{[k]}a_{i}^{\dag }a_{i+1}+h.c,
\end{equation}%
where $a_{i}^{\dag },a_{i}$ are the fermion operators. This describes a
simple hopping process in the lattice.

According to our SPMC, we can present different \ models (labelled
by different positive integer $k(\in \{0,1,2,...\})$) through
pre-engineering of the coupling strength as

\begin{equation}
J_{i}=J_{i}^{[k]}=\sqrt{i\left( N-i\right) }
\end{equation}%
for even $i$ \ and
\[
J_{i}=J_{i}^{[k]}=\sqrt{\left( i+2k\right) \left( N-i+2k\right) }
\]%
for odd $i$. By a straightforward calculation, one can find the
k-dependent spectrum%
\begin{equation}
\varepsilon _{n}=-N+2(n-k)-1
\end{equation}%
for $n=1,2,...,N/2,$and
\begin{equation}
\varepsilon _{n}=-N+2(n+k)-1
\end{equation}%
for $n=N/2+1,...,N$ . The corresponding k-dependent eigenstates are

\begin{equation}
\left\vert \phi _{n}\right\rangle =%
\mathop{\textstyle\sum}%
_{i=1}^{N}c_{ni}\left\vert i\right\rangle =%
\mathop{\textstyle\sum}%
_{i=1}^{N}c_{ni}a_{i}^{\dag }\left\vert 0\right\rangle
\end{equation}%
where the coefficients can be determined by

\begin{eqnarray}
c_{n2} &=&\frac{\varepsilon _{n}c_{n1}}{\sqrt{\left( 1+2k\right) \left(
N-1+2k\right) }}, \\
&&....  \nonumber \\
0\text{ } &=&\sqrt{\left( i+2k+1\right) \left( N-i+2k-1\right) }%
c_{ni+2}-\varepsilon _{n}c_{ni+1}  \nonumber \\
&&+\sqrt{i\left( N-i\right) }c_{ni}\text{\ \ \ (}i\text{ is even)},
\nonumber \\
0\text{ } &=&\sqrt{\left( i+2k\right) \left( N-i+2k\right) }%
c_{ni}-\varepsilon _{n}c_{ni+1}  \nonumber \\
&&+\sqrt{\left( i+1\right) \left( N-i-1\right) }c_{ni+2}\text{\ \ (}i\text{
is odd),}  \nonumber \\
&&....  \nonumber \\
c_{nN} &=&\frac{\sqrt{\left( 1+2k\right) \left( N-1+2k\right) }c_{nN-1}}{%
\varepsilon _{n}}.  \nonumber
\end{eqnarray}%
It is obvious that the model proposed in ref. \cite{Matt} is just
the special case of our general model in $k=0$. For arbitrary $k$,
we can easily check that it meets the our SPMC by a
straightforward calculation. Thus we can conclude that these spin
systems with a set $S^{[k]}$ of pre-engineered couplings
$J_{i}^{[k]}$ can serve the perfect quantum channels allow to
transfer the quantum information of spin qubits.

In the above arguments we show the possibibility \ to implemented
the perfect state transfer of any quantum state over arbitrarily
long distances in a quantum spin chain. The crucial point to this
end is that one need to locally engineer couplings between the
spins in the specific way. Now we need consider the possibility to
engineer couplings for the practical quantum information
processing. In usual, we expect to transfer quantum information
over much longer distance. For this purpose the spin chain must be
longer and thus contain too many degrees of freedom. Since the
dimension of the Hilbert space of the many-body system grows
exponentially with the size of the system, there must be enormous
parameters to be exactly engineered. In this sense it is almost
impossible to engineer a real spin system so that it possesses
energy levels to exactly satisfy the SPMC. To overcome the
difficulties, there is a naive way that one choose some special
states to be transported, which is a coherent superposition of
commensurate part of the whole set of eigenstates. For example, we
consider a truncated Gaussian wavepacket for an anharmonic
oscillator with lower eigenstates to be harmonic. It is obvious
that such system allows some special states to transfer with high
fidelity. We can implement such approximate harmonic system in  a
natural spin chain without  the pre-engineering of  couplings. Our
strategy is to apply a modulated external field.

Let us consider the Hamiltonian of $(2N+1)$-site spin-$\frac{1}{2}$
ferromagnetic Heisenberg chain

\begin{equation}
H=-J\sum_{i=1}^{2N}\overrightarrow{S}_{i}\cdot \overrightarrow{S}%
_{i+1}+\sum_{i=1}^{2N+1}B(i)S_{i}^{z}
\end{equation}%
with the uniform coupling strength $-J<0,$ but in the parabolic
magnetic field

\begin{equation}
B(i)=2B_{0}(i-N-1)^{2}
\end{equation}%
where $B_{0}$ is a constant. In single-excitation invariant
subspace with the fixed z-component of total spin
$S^{z}=\frac{2N-1}{2}$, this model is equivalent to the spin-less
fermion hopping model with the Hamiltonian

\begin{equation}
H=-\frac{J}{2}\sum\limits_{i=1}^{2N}(a_{i}^{\dag }a_{i+1}+h.c)+\frac{1}{2}%
\sum_{i=1}^{2N+1}B(i)a_{i}^{\dag }a_{i}
\end{equation}
where we have neglected a constant in the Hamiltonian for
simplicity. For the single-particle case with the site basis
\[
\{\stackrel{\qquad \ \ \ \ \ \ \ n^{\prime }th}{\left\vert
n\right\rangle =\left\vert 0,0,...,1,0...\right\rangle
}|n=1,2,..\}
\]%
the matrix presentation of the Hamiltonian (16) is%
\begin{eqnarray}
H_{JJ} &=&\sum\limits_{n}\{-\frac{1}{2}E_{J}(\left\vert n+1\right\rangle
\left\langle n\right\vert +\left\vert n+1\right\rangle \left\langle
n\right\vert ) \\
&&+4E_{c}(n-n_{g})^{2}\left\vert n\right\rangle \left\langle n\right\vert \}
\nonumber
\end{eqnarray}%
which is just the same as that of the Hamiltonian of Josephson
junction in the Cooper-pair number basis Ref.\cite{Shn} for
$E_{J}=J,E_{c}=2B_{0}$. Analytical analysis and numerical results
can show that the lower energy spectrum is indeed  quasi-harmonic
in the case $E_{J}\gg E_{c}$\cite{Shi}. Although the eigenstates
of the Hamiltonian (14) does not satisfy the SPMC precisely,
especially for high energy range, there must exist some Gaussian
wavepacket states expanded by the lower eigenstates, which can be
transferred near perfectly.

\vspace*{-3.5cm}
\begin{figure}[tbp]
\hspace{24pt}\includegraphics[width=8cm,height=12cm]{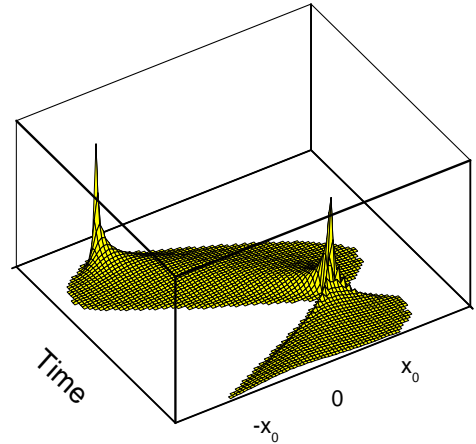}
\vspace*{-2.0cm} \caption{Schematic illustration of the time
evolution of a Gaussian wavepacket. It shows that the near-perfect
state transfer over a long distance is possible in the
quasi-harmonic system.}
\end{figure}
\vspace*{-0.0cm}

We consider a Gaussian wavepacket at $t=0$, $x=N_{A}$ as the initial state
\begin{equation}
\left\vert \psi (N_{A},0)\right\rangle =C\sum_{i=1}^{2N+1}e^{-\frac{1}{2}%
\alpha ^{2}(i-N_{A}-1)^{2}}\left\vert i\right\rangle
\end{equation}
where $\left\vert i\right\rangle $ denotes the state with $2N$ spins in down
state and only $i$th spin in up state, $C$ is the renormalization factor.
The coefficient $\alpha ^{2}$ is determined by the width of the Gaussian
wavepacket $\Delta$
\begin{eqnarray}
\alpha ^{2} =\frac{4\ln 2}{\Delta^{2}}.
\end{eqnarray}
The state $\left\vert \psi (0)\right\rangle $ evolves to

\begin{equation}
\left\vert \psi (t)\right\rangle =e^{-iHt}\left\vert \psi
(N_{A},0)\right\rangle
\end{equation}%
at time $t$ and the fidelity of state $\left\vert \psi (0)\right\rangle $
transferring to the position $N_{B}$ is defined as
\begin{equation}
F(t)=\left\vert \left\langle \psi (N_{B},0)\right\vert e^{-iHt}\left\vert
\psi (N_{A},0)\right\rangle \right\vert .
\end{equation}
In Fig.1 the evolution of the state $\left\vert \psi
(0)\right\rangle $ is illustrated schematically. From the
investigation of Ref.\cite{Shi}, we know
that for small $N_{A}=-N_{B}=-x_{0}$, where $N_{B}$ is the mirror counterpart of $%
N_{A}$, but in large $\Delta$ limit, if we take $B_{0}=8\left( \frac{\ln 2}{%
\Delta^{2}}\right) ^{2}$, $F(t)$ has the form
\begin{equation}
F(t)=\exp [-\frac{1}{2}\alpha ^{2}N_{A}^{2}(1+cos\frac{2t}{\alpha
^{2}})]
\end{equation}%
\vspace*{-1.0cm}
\begin{figure}
\hspace{24pt}\includegraphics[width=6cm,height=9cm]{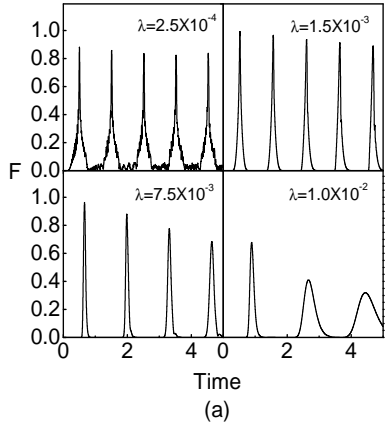}
\vspace*{-3.0cm}

\hspace{24pt}\includegraphics[width=6cm,height=9cm]{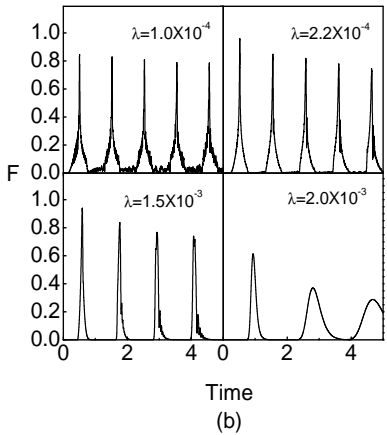}
\vspace*{-3.0cm}

\hspace{24pt}\includegraphics[width=6cm,height=9cm]{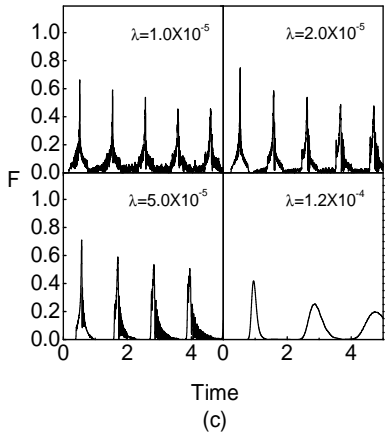}
 \vspace*{%
-1.5cm}

\caption{The fidelities $F(t)$ of the transmission of the Gaussian
wavepackets with $\Delta=2,4$ and $6$ over the distance $L=500$
are plotted for different values of $\lambda$ in Fig.2(a), (b) and
(c). It shows that there exist the optimal values of $\lambda$ to
get high fidelities up to $0.748,0.958$ and $0.992$,
respectively.}
\end{figure}

which is a periodic function of $t$ with maxima 1. This is in
agreement with our above analysis. However, in quantum
communication, what we concern is the behavior of $F(t)$ in the
case of the transfer distance $L\gg \Delta $, where $L=2\left\vert
N_{A}\right\vert =2\left\vert N_{B}\right\vert $. For this purpose
the numerical method is performed for the case $L=500,\Delta
=2,4,6$ and $B_{0}=8\left( \frac{\ln 2}{\Delta ^{2}}\right)
^{2}\lambda $. The factor $\lambda $ determines the maximum
fidelity and then the optimal
field distribution can be obtained numerically. In Fig.2(a), (b) and (c) the functions $%
F(t)$ are plotted for different values of $\lambda $. It shows that for the
given wavepackets with $\Delta =2,4$ and $6$, there exists a range of $%
\lambda $, during which the fidelities $F(t)$ are up to $0.748,0.958$ and $%
0.992$ respectively. For finite distance, the maximum fidelity decreases as
the width of Gaussian wavepacket increases. On the other hand, the strength
of the external field also determines the value of the optimal fidelity for
a given wavepacket. Numerical results indicate that it is possible to
realize near-perfect quantum state transfer over a longer distance in a
practical ferromagnetic spin chain system.

In summary, we have shown that a perfect quantum transmission can
be realized through a universal quantum channel provided by a
quantum spin system with spectrum structure, in which each
eigenenergy is commensurate and matches with the corresponding
parity. According to this SPMC for the a mirror inversion symmetry
\cite{MIS}, we can implement the perfect quantum information
transmission with several novel pre-engineered quantum spin
chains. For more practical purpose, we prove that an approximately
commensurate spin system can also realize near-perfect quantum
state transfer in a ferromagnetic Heisenberg chain with uniform
coupling constant, but in an external field. Numerical method has
performed to study the fidelity for the system in a parabolic
magnetic field. The external field plays a crucial role in the
scheme. It induces a lower quasi-harmonic spectrum, which can
drive a Gaussian wavepacket from the initial position to its
mirror counterpart. The fidelity depends on the initial position
(or distance $L$), the width of the wavepacket $\Delta $ and the
magnetic field distribution $B(i)$ via the factor $\lambda $. Thus
for given $L$ and $\Delta $, proper choice of the factor $\lambda
$ can achieve the optimal fidelity. Finally, we conclude that it
is possible to implement near-perfect Gaussian wavepacket
transmission over a longer distance in many-body system.
\newline

This work of SZ is supported by the Innovation Foundation of Nankai
university. CPS also acknowledge the support of the CNSF (grant No.
90203018) , the Knowledge Innovation Program (KIP) of the Chinese Academy of
Sciences, the National Fundamental Research Program of China (No.
001GB309310).

\end{document}